\pdfoutput=1

\documentclass[10pt,twocolumn,letterpaper]{article}

\usepackage{wacv}              %

\usepackage{array}
\usepackage{xurl}
\usepackage{amsfonts}
\usepackage{amsthm}
\usepackage{textcomp}
\usepackage{marvosym}
\usepackage{microtype}
\newcommand{\MCD}{\mathrm{MCD}}
\definecolor{wacvblue}{rgb}{0.21,0.49,0.74}
\usepackage[pagebackref,breaklinks,colorlinks,allcolors=wacvblue]{hyperref}
\AddToShipoutPicture*{\AtPageLowerLeft{\put(\LenToUnit{0.8125in},\LenToUnit{0.32in}){\parbox{6.875in}{\scriptsize\textcopyright~2027 IEEE. Personal use of this material is permitted. Permission from IEEE must be obtained for all other uses, in any current or future media, including reprinting/republishing this material for advertising or promotional purposes, creating new collective works, for resale or redistribution to servers or lists, or reuse of any copyrighted component of this work in other works.}}}}

\hypersetup{pdftitle={Mutual Colormap Dependence: How Recoloring Biomedical Images Changes Vision-Language Model Answers},
  pdfauthor={Siddharth Vohra, Runmin Jiang, Min Xu}}

\def\wacvPaperID{*****} %
\def\confName{WACV}
\def\confYear{2027}

\title{Mutual Colormap Dependence: How Recoloring Biomedical Images
Changes Vision-Language Model Answers}

\author{Siddharth Vohra\textsuperscript{1,2} \qquad Runmin Jiang\textsuperscript{1} \qquad Min Xu\textsuperscript{1,\Letter}\\
\textsuperscript{1}Carnegie Mellon University, Pittsburgh, PA, USA\\
\textsuperscript{2}Amazon Web Services AI Native, Pittsburgh, PA, USA\\
{\tt\small \{svohra, runminj, mxu1\}@andrew.cmu.edu}
}

\begin{document}
\maketitle
\begin{NoHyper}\let\thefootnote\relax\footnotetext{Work done by Siddharth Vohra does not relate to the position he currently holds at Amazon Web Services AI Native.\quad\textsuperscript{\Letter}Corresponding author.}\footnotetext{This study used only public datasets with no identifiable patient data.}\end{NoHyper}
\begin{abstract}
Vision-language models (VLMs) are being built to answer questions about
biomedical images, yet they never see the measured intensities, only the colors
that a colormap has assigned to them. For questions about image content, the
colormap should not change the answer. We call the extent to which it does
Mutual Colormap Dependence (MCD), the mutual information between the colormap
and a model's answer, which is zero exactly when every colormap yields the same
answer distribution. Because VLMs sample their answers, we compare
recoloring with a control that uses the same number of calls to ask again
about the unchanged gray image, so that ordinary resampling is not mistaken for
an effect of the colormap. We test GPT-5.4, Gemini 3.1 Pro, Claude Sonnet 4.6, and
Qwen3-VL-235B under six luminance-monotonic colormaps (gray, viridis, plasma,
inferno, magma, and cividis) on 50 medical visual-question-answering items from
16 SLAKE images, 64 ChestX-ray14 chest radiographs, and 200 BBBC039
fluorescence-microscopy images for exact nucleus counting. Every prompt tells
the model to ignore the colormap. Even so, with three answers per colormap on
the chest radiographs, the answers of all four models depend on the colormap
(all Holm-adjusted $p\leq 0.002$). For Claude, some colored map
overturns the majority answer on 29 of the 64 radiographs, where resampling
gray alone overturns about 4, and Qwen's accuracy falls from 0.54 under gray to
between 0.32 and 0.41 under the colored maps, mostly because it abstains. With one answer per colormap,
recoloring adds disagreement beyond repeated sampling in 5 of 12 model-panel
cells under per-cell 95\% intervals, and repeating the instruction to ignore
color does not reliably reduce it. The colormap is therefore part of what a
VLM sees, and it can change the answer.
\end{abstract}

\begin{figure*}[t]
\centering
\setlength{\tabcolsep}{1.6pt}
\renewcommand{\arraystretch}{1.1}
\newcommand{\teaserimg}[1]{\includegraphics[width=0.158\textwidth]{example_#1.jpg}}
\begin{tabular}{cccccc}
\small gray & \small viridis & \small plasma & \small inferno & \small magma & \small cividis\\
\teaserimg{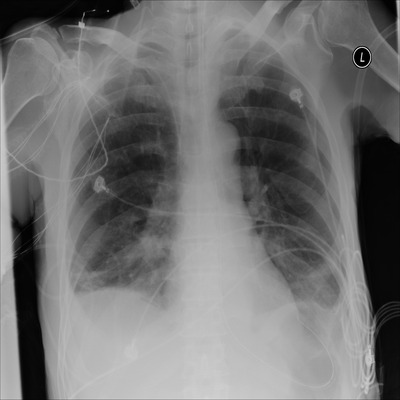} & \teaserimg{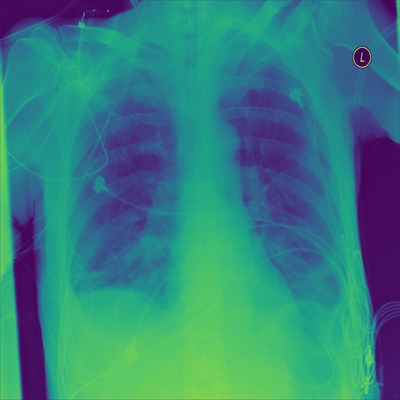} & \teaserimg{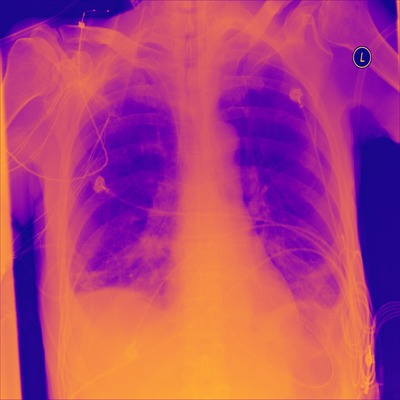} &
\teaserimg{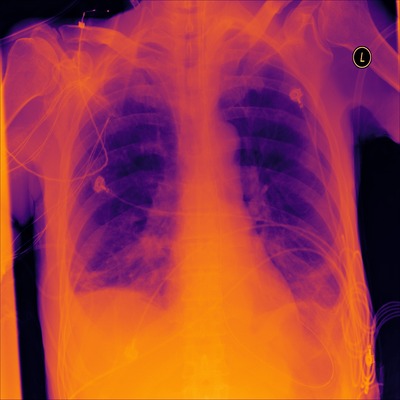} & \teaserimg{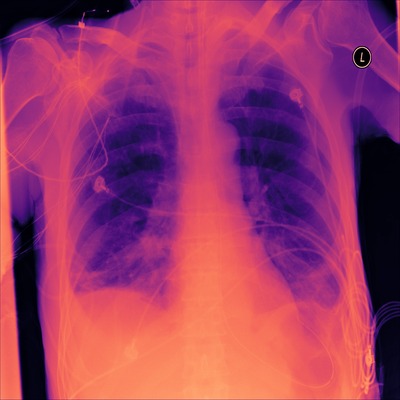} & \teaserimg{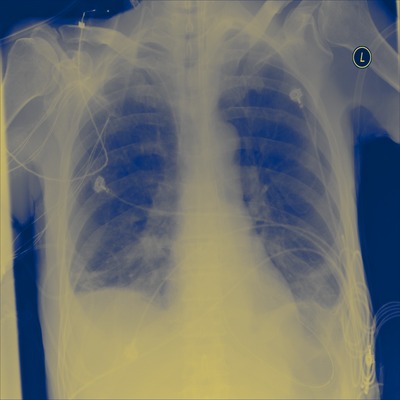}\\
\small yes\enspace yes\enspace yes & \small no\enspace no\enspace no &
\small yes\enspace yes\enspace yes & \small no\enspace no\enspace no &
\small no\enspace no\enspace no & \small yes\enspace yes\enspace yes\\
\end{tabular}

\vspace{2pt}
{\small The gray image asked six more times:\enspace
yes\enspace yes\enspace yes\enspace yes\enspace yes\enspace yes}
\vspace{1pt}
\caption{\textbf{Recoloring changes the answer, while asking again about the
gray image does not.} Claude Sonnet 4.6's answers to ``Is there evidence of
atelectasis in this chest X-ray?'' for one ChestX-ray14 radiograph labeled
positive, three under each colormap and six more for the gray image.}
\label{fig:teaser}
\end{figure*}

\section{Introduction}\label{sec:intro}

A chest radiograph or a fluorescence micrograph begins as a grid of measured
intensities, and a colormap turns these numbers into colors before anyone looks
at the image. The DICOM standard specifies a grayscale display function and how to
calibrate a display to it~\cite{dicom2026gsdf}, and the same standard defines pseudo-color
palettes such as Hot Iron, which it notes is often used in nuclear
medicine~\cite{dicom2026palettes}. Matplotlib, a widely used scientific plotting
library, offers perceptually uniform maps such as viridis and magma and renders
a single-channel image in viridis unless told
otherwise~\cite{matplotlib_colormaps,matplotlib_default}. The choice matters to
human readers. A poorly chosen colormap, such as a rainbow, can add apparent
features that are not in the measured data and hide ones that
are~\cite{borland2007rainbow,kovesi2015good,crameri2020misuse}, and
colorizing mammograms while keeping a calibrated luminance scale can change how
well observers detect lesions~\cite{knippa2020colorized}.

Vision-language models (VLMs) now read these images as well. A clinician or
scientist can paste a rendered figure or a screenshot into a general-purpose
model and ask about it. The model receives only the colored pixels, so the
lookup table is part of its input: a radiograph plotted with Matplotlib's
defaults arrives in viridis, and a fluorescence image is often shown in
pseudocolor. A model that answers yes for a radiograph shown in gray and no for
the same radiograph shown in magma is letting a display setting decide its
answer. Figure~\ref{fig:teaser} shows that this happens.

Measuring this dependence raises two problems. Accuracy is a blunt instrument,
because a model can move from one wrong answer to another, or gain on some
images what it loses on others, with no change in accuracy. These models also
sample their answers. Asking the same question twice about the same image can
already produce two answers, so a disagreement between a gray and a magma
rendering may be nothing more than a second draw. Existing benchmarks that test
robustness to recoloring or colormaps score accuracy on the transformed
inputs~\cite{liang2025colorbench,shrivastava2026thermeval}, so they see neither
changes between wrong answers nor the difference between a transformation and a
resample.

We define Mutual Colormap Dependence (MCD) as the conditional mutual
information $I(C;A\mid I,q)$ between a uniformly drawn colormap $C$ and the
model's sampled answer $A$, for a fixed scalar image $I$ and question $q$. MCD
is zero exactly when every colormap gives the same answer distribution, it
registers every change, including a change from one wrong answer to another,
and it needs no reference labels. Colormaps differ in hue, in how luminance
rises with intensity, and in local contrast, so MCD measures dependence on the
display mapping as a whole, which is exactly what a user changes when choosing
a colormap. We measure it in two steps. A matched screen
asks the question once under each of six colormaps and compares the
disagreement among the six answers, which we call orbit drift, with the
disagreement among six answers to the gray image, which we call same-render
drift. Both arms use the same number of calls, so their difference shows
whether recoloring adds disagreement beyond what asking again already
produces. Where more calls are possible, three answers per colormap estimate
MCD directly, and a permutation test decides whether the answer depends on the
colormap.

We test four frontier VLMs from four developers, GPT-5.4, Gemini 3.1 Pro,
Claude Sonnet 4.6, and the open-weight Qwen3-VL-235B. The three panels cover
both display conventions: SLAKE medical images and ChestX-ray14 chest
radiographs, which are normally read in gray, and BBBC039 fluorescence
micrographs, which are often shown in pseudocolor. All six colormaps increase in
luminance with intensity and are perceptually ordered maps of the kind
visualization research recommends~\cite{kovesi2015good,crameri2020misuse}, so
none adds the false boundaries of a rainbow map, and any dependence we find
cannot be blamed on a poorly designed palette. Every prompt also tells the model to ignore the colormap, so any
dependence we find persists despite an explicit instruction.

With three answers per colormap on the 64 chest radiographs, every model shows
a colormap-answer association after Holm correction. For Claude,
some colored map overturns the majority answer on 29 of the 64 radiographs,
where resampling gray alone overturns about 4, and 17 of these 29 are among the
37 whose gray majority matches the label. Qwen abstains so often under color
that its accuracy falls by up to 22 points. With one answer per colormap, recoloring adds disagreement beyond
repeated sampling in five of twelve model-panel cells, including every Claude
panel. Repeating the instruction to ignore color does not reliably reduce the
disagreement. We make three contributions.
\begin{itemize}
\item Mutual Colormap Dependence (MCD), an information-theoretic measure of how
much a VLM's answer distribution changes with the colormap, which is zero
exactly when the model is colormap-invariant.
\item A screen-and-confirm protocol that compares recoloring with resampling
at an equal number of calls, consisting of a comparison of orbit drift with
same-render drift at six calls each and a three-answer permutation test of MCD.
\item Evidence that four frontier VLMs violate colormap invariance on chest
radiographs despite an explicit instruction to ignore color, with consequences
for accuracy and abstention, together with a release of the prompts, all 20,880
answers, and the analysis code from which every result in the paper can be
rebuilt.
\end{itemize}

\section{Related Work}\label{sec:related}

\subsection{Colormaps and Robustness}

Visualization research has long studied how the colormap changes the way
people read scalar data. Perceptual ordering, rainbow artifacts, diverging
maps, and plotting-tool defaults all affect readability~\cite{ware1988color,
rogowitz1998rainbow,borland2007rainbow,moreland2009diverging,kovesi2015good,
matplotlib_colormaps,matplotlib_default,crameri2020misuse}, and observers infer
quantity from lightness and apparent opacity in ways that depend on the
background color~\cite{schloss2019mapping}. Vision models can rely on texture and other
cues outside the intended signal and can fail under common
corruptions~\cite{hendrycks2019robustness,geirhos2019texture,
geirhos2020shortcut}. Behavioral and metamorphic testing turn a property that
should hold into a controlled change of the input~\cite{chen2018metamorphic,
ribeiro2020checklist}. Visual question answering models can lean on
language priors~\cite{goyal2017vqa,agrawal2018dont}, which motivated
benchmarks that control for them~\cite{hudson2019gqa}, and VLMs can miss
low-level visual detail~\cite{rahmanzadehgervi2024vlms}. Recent multimodal studies bring color into this
picture. ColorBench measures whether VLM performance stays consistent under
color transformations~\cite{liang2025colorbench}, ThermEval finds that VLM
robustness to colormap transformations of thermal images is
inconsistent~\cite{shrivastava2026thermeval}, and FUGU finds evidence that
some VLM failures on synthetic scatter plots arise in the handoff between the vision and
language components~\cite{tartaglini2025fugu}. These studies measure task
performance. We hold the scalar image and the question fixed, measure how much
the answer distribution itself depends on the display, and compare that
dependence with the variation a stochastic model shows when it is simply asked
again.

\subsection{Uncertainty from Disagreement}

Disagreement among several answers is a familiar uncertainty signal. Selective
classification studies the tradeoff between how many predictions a model makes
and how many of them are wrong~\cite{elyaniv2010foundations,
geifman2017selective}. Modern networks can be poorly
calibrated~\cite{guo2017calibration}, their uncertainty can degrade under
distribution shift~\cite{ovadia2019uncertainty}, and ensembles are one
response~\cite{lakshminarayanan2017deep}. For language models,
self-consistency samples several answers to the same
input~\cite{wang2023selfconsistency}, and semantic entropy measures how much
sampled answers disagree in meaning~\cite{kuhn2023semantic,
farquhar2024semantic}. Test-time augmentation instead perturbs the input and
reads uncertainty from the spread of the predictions, including in medical
image analysis~\cite{wang2019tta}. MCD is the same decomposition, total entropy
minus expected entropy, that BALD applies over model
parameters~\cite{houlsby2011bald}, with the expectation taken over displays. Our orbit is
test-time augmentation over the display, and our same-render control is
self-consistency with the same number of calls. Traub
\emph{et al.}~\cite{traub2024selectiveclassificationflaws} identify
shortcomings of multi-threshold metrics, including the area under the
risk-coverage curve (AURC), which over-weights confident failures. We use AURC only to
compare two scores on identical predictions, with paired intervals.

\subsection{Medical VLMs and Evaluation Panels}

Medical systems such as LLaVA-Med and Med-Flamingo adapt VLMs to biomedical
images~\cite{li2023llavamed,moor2023medflamingo}. Our test needs only query
access, so it applies to such systems unchanged, and we focus on the
general-purpose frontier models that clinicians and scientists can query
directly. Our panels come from three public datasets: the SLAKE medical
visual-question-answering benchmark~\cite{liu2021slake}, ChestX-ray14 with
labels extracted by natural language processing
(NLP)~\cite{wang2017chestxray14,nihchestxray}, and BBBC039 with manually
annotated nucleus masks~\cite{bbbc039,caicedo2019evaluation}.

\section{Measuring Colormap Dependence}\label{sec:method}

\begin{table}[t]
\caption{\textbf{Quantities at a Glance}}
\label{tab:quantities}
\centering
\footnotesize
\setlength{\tabcolsep}{3pt}
\begin{tabular}{@{}lc>{\raggedright\arraybackslash}p{0.5\columnwidth}@{}}
\toprule
Quantity & Calls & Role\\
\midrule
MCD, Eq.~\eqref{eq:mcd} & target & Effect of the colormap on answers\\
Plug-in MI, Eq.~\eqref{eq:multirepeat-mcd} & 18 & Mutual information (MI) estimate, permutation-tested\\
Orbit drift, Eq.~\eqref{eq:orbit-drift} & 6 & Disagreement across colormaps\\
Same-render drift, Eq.~\eqref{eq:same-drift} & 6 & Disagreement across gray repeats\\
$\Delta\widehat D$ & 12 & Disagreement added by recoloring\\
Gray leverage & 6 & Gray versus a typical chromatic map\\
$\Delta$AURC & 12 & Recoloring versus resampling for uncertainty\\
\bottomrule
\end{tabular}

\smallskip
\parbox{\columnwidth}{\footnotesize Calls are per item. MCD is the population
target, and the other quantities are computed from observed answers.}
\end{table}

\subsection{Setting}

We hold a scalar image $I$ and a question $q$ fixed and vary only the lookup
table that renders the image. Let \mbox{$C=\{c_1,\ldots,c_K\}$} be a set of
deterministic lookup tables, so that the model sees the rendered image
$c_k(I)$. We restrict $q$ to questions about anatomy, structure, relative
intensity, or a count, whose correct answer is the same under every rendering,
and we exclude questions about color. For these questions colormap invariance
is a property the model should have, and any departure from it is a behavior
worth measuring. The measurement uses only the model's answers, so it needs
only query access to the model. Table~\ref{tab:quantities} summarizes the
quantities defined below and how many calls each needs.

\subsection{Mutual Colormap Dependence}

MCD measures how much the model's answer distribution changes with the
colormap. Let
\begin{align*}
p_k(a)&=P_M(A=a\mid c_k(I),q)
\quad\text{and}\\
\bar p(a)&=\frac{1}{K}\sum_{k=1}^{K}p_k(a).
\end{align*}
For a colormap drawn uniformly from $C$, define
\begin{equation}\label{eq:mcd}
\begin{split}
\MCD(I,q)
&=H(\bar p)-\frac{1}{K}\sum_{k=1}^{K}H(p_k)\\
&=I(C;A\mid I,q).
\end{split}
\end{equation}
The first term is the variability of the answer when the colormap is free, and
the second is its variability when the colormap is fixed, so MCD is the part of
the answer's variability that the colormap explains. Equation~\eqref{eq:mcd} is
the generalized Jensen-Shannon divergence of the $p_k$ with equal
weights~\cite{lin1991divergence}. Because entropy is strictly concave, MCD is
non-negative and is zero if and only if all $p_k$ are equal, and because mutual
information cannot exceed $H(C)$, it is at most $\log K$. MCD registers every
change in the answer distribution, whether the model moves between yes and no,
between two wrong answers, or between answering and declining. A reply flagged as
an abstention counts as one answer, whatever else it holds, and an unflagged
\texttt{unknown} counts as another. Each is an answer like any other, since a
model that answers under gray and declines under magma has given the user a
different answer.

\subsection{Estimating MCD from Repeated Answers}\label{sec:mcd-estimate}

With $R$ separately issued answers under every colormap, we estimate each $p_k$
by the empirical answer frequencies $\widehat p_k$ and compute the plug-in
estimate
\begin{equation}\label{eq:multirepeat-mcd}
\begin{split}
\widehat I_{\mathrm{pi}}(C;A\mid I,q)
={}&H_{\mathrm{pi}}\left(\frac{1}{K}\sum_{k=1}^{K}\widehat p_k\right)\\
&-\frac{1}{K}\sum_{k=1}^{K}H_{\mathrm{pi}}(\widehat p_k),
\end{split}
\end{equation}
where $H_{\mathrm{pi}}$ is plug-in entropy in nats. The subtracted term, the
within-map entropy, shows how much the model varies when the display is fixed.
Plug-in mutual information is biased upward at small $R$, so we decide
dependence with a permutation test. We shuffle the colormap labels within each
item and repeat block, which keeps the answers and the repeat counts, and we
apply Holm correction across the four models. The mean of the permutation null
also calibrates the bias, so we report the plug-in estimate minus that mean as
a descriptive effect size and base inference on the test itself.

\subsection{A Matched Screen}\label{sec:screen}

Three answers under each of six maps take 18 calls per item. For a lighter
screen we ask once under each map and summarize the six answers $\hat a_k$ by
their plug-in entropy, the orbit drift
\begin{equation}\label{eq:orbit-drift}
\widehat D_{\mathrm{orbit}}(I,q)
=H_{\mathrm{pi}}\left(
\frac{1}{K}\sum_{k=1}^{K}\delta_{\hat a_k}
\right),
\end{equation}
where $\delta_{\hat a_k}$ is a point mass at the observed answer. On this scale,
six identical answers give 0, a five-to-one split gives 0.451, a three-to-three
split gives $\log 2=0.693$, and six different answers give $\log 6=1.792$ nats.
One answer per map cannot identify the $p_k$, because a model that samples
freely can disagree with itself even when every colormap yields the same
distribution. We therefore also query the gray rendering of each item $K=6$
times with the same prompt and settings, and compute the same-render drift
\begin{equation}\label{eq:same-drift}
\widehat D_{\mathrm{same}}(I,q)
=H_{\mathrm{pi}}\left(
\frac{1}{K}\sum_{r=0}^{K-1}\delta_{\hat a^{\,g}_r}
\right),
\end{equation}
where $\hat a^{\,g}_r$ is the answer to gray repeat $r$. Gray is the standard
display for radiographs, CT, and MRI and the plain view of any single-channel
image, so this arm measures how much the model varies under a fixed reference
view. The paired contrast
\begin{equation*}
\Delta\widehat D(I,q)=\widehat D_{\mathrm{orbit}}(I,q)-\widehat D_{\mathrm{same}}(I,q)
\end{equation*}
uses six calls in each arm. A positive mean contrast means that the six
colormaps produced more disagreement than six repeats of gray on the evaluated
items. The contrast is a screening statistic, and the permutation test of
Section~\ref{sec:mcd-estimate} is the confirmatory step.
Under invariance every map shares gray's answer distribution, so both arms
are six draws from the same distribution and the contrast has mean zero. Away
from invariance, a difference in noise between gray and the colored maps also
moves the contrast, which is why it does not estimate MCD. The screen therefore
tests for excess disagreement over the reference view, the permutation test
tests for any change across maps, and a cell can be flagged by one and not the
other.

\subsection{Leaving Out Gray}

Gray is the only achromatic map, so a large orbit drift could in principle come
from a single split between gray and the colored maps. We recompute orbit drift
with each map left out in turn, which gives six five-map orbits of equal size.
With $\widehat D_{-g}$ the score without gray, gray leverage is
$\overline{\widehat D}_{-\mathrm{chrom}}-\widehat D_{-g}$, where the first term
is the mean score after omitting each chromatic map in turn. A positive value
means that gray contributes more disagreement than a typical chromatic map.

\subsection{Disagreement as an Error Signal}\label{sec:selective}

Disagreement can also tell a user which answers to trust. Both signals score
the same prediction, the gray answer from the first repeat, r0. Orbit
uncertainty uses that answer and the five chromatic answers, and same-render
uncertainty uses that answer and five of the six gray repeats, so each signal
uses six calls. We rank answers by each signal and compute AURC, counting
abstentions as errors, and items with tied uncertainty enter the risk-coverage
curve as one block. The primary same-render score leaves out gray repeat r5.
We evaluate all six choices and count an advantage for orbit uncertainty only
when the paired AURC interval favors it and every choice preserves the sign.
As a reference we also score the model's stated confidence, which needs one
call.

\section{Experimental Setup}\label{sec:setup}

\subsection{Panels and Questions}

\textbf{SLAKE}~\cite{liu2021slake} contributes 50 English closed-answer
questions about 16 CT, MRI, and X-ray images, 42 of them yes/no questions with
21 yes and 21 no answers, such as ``Does the picture contain liver?'', and 8 of
them organ-choice questions, such as ``Which organ is abnormal, heart or
lung?'' Answers are restricted to yes, no, heart, lung, and unknown. Several questions share an image, so the image is the sampling unit.

\textbf{ChestX-ray14}~\cite{wang2017chestxray14,nihchestxray} contributes 64
radiographs from 63 patients and four questions, about atelectasis, opacity, pneumonia, and
pneumothorax, each asked of eight positive and eight negative cases under the
public NLP-derived labels, where a negative case lacks the finding's label but
may carry others. Each radiograph carries one question of the form
``Is there evidence of atelectasis in this chest X-ray?'' ChestX-ray14 has no
opacity label, so the opacity cases are drawn and scored with the Effusion
label, a finding that appears on the radiograph as a pleural opacity, so an
opacity negative can still show another kind of opacity.
Accuracy and AURC count an answer as correct when it matches the label, so they
inherit its noise~\cite{majkowska2020chest}, while MCD, drift, and changes of the majority answer are
computed from the answers alone.

\textbf{BBBC039}~\cite{bbbc039,caicedo2019evaluation}, from the Broad Bioimage
Benchmark Collection~\cite{ljosa2012bbbc}, contributes 200 fluorescence images
of nuclei, with reference counts derived from the
manually annotated masks. The model must return the exact number of nuclei, so
this panel tests whether a precise output stays fixed under recoloring, a
stricter standard than approximate counting. All three datasets are
public: BBBC039 is released under CC0~\cite{bbbc039}, the SLAKE annotations
under CC BY 4.0~\cite{slake_card}, and ChestX-ray14 by the NIH Clinical Center
for unrestricted use with attribution~\cite{nihchestxray}.

\subsection{Colormaps and Rendering}

The six maps are gray, viridis, plasma, inferno, magma, and cividis. Each
source image is read as one scalar channel, downscaled only when its long edge
exceeds 1024 pixels, min-max normalized, passed through a fixed 256-entry
lookup table, and saved as an 8-bit RGB PNG. The 16-bit BBBC039 images are
first converted to 8 bits by clipping at 255, which saturates pixels in 197
of the 200 images. Across the panel the median image has 24\% of its pixels
saturated, and the most affected image has 45\%. Clipping and min-max
normalization precede the lookup table, so all six renderings and both arms
receive the same scalar field, and the BBBC039 results describe these clipped
images. All six maps increase in luminance with
intensity. After 8-bit quantization each keeps a Spearman correlation above
0.9999 between intensity and Rec.~709 luma, with at most six reversals between
adjacent entries, none larger than 0.002 in luma, and only viridis and cividis
repeat a color, in two and three entries.

\subsection{Prompts}\label{sec:prompt}

Every system prompt tells the model to ignore the colormap. The medical prompts
read ``Ignore the colormap or hue; only anatomy, structure, texture, and
relative intensity patterns matter,'' and the BBBC039 prompt names spatial
structure, boundaries, and relative intensities instead. The prompts allow the
model to abstain when the answer cannot be decided from the image and ask for a
JSON object with an answer, or a count for BBBC039, an abstention flag, and a
confidence. A separate arm, the
\textbf{reinforced invariance reminder}, keeps this prompt and appends one
user-level sentence:
\begin{quote}
\emph{The image may be pseudocolored arbitrarily; treat the chosen colormap as
presentation only and not as evidence.}
\end{quote}

\begin{table}[t]
\caption{\textbf{Design of the Experiment}}
\label{tab:validation-matrix-design}
\centering
\footnotesize
\setlength{\tabcolsep}{2.5pt}
\begin{tabular}{llrrrr}
\toprule
Panel & condition & rows & images & maps & calls per row \\
\midrule
SLAKE & six-map raw & 50 & 16 & 6 & 6 \\
SLAKE & six-repeat gray & 50 & 16 & 1 & 6 \\
SLAKE & reinforced reminder & 50 & 16 & 6 & 6 \\
ChestX-ray14 & repeated six-map & 64 & 64 & 6 & 18 \\
ChestX-ray14 & six-repeat gray & 64 & 64 & 1 & 6 \\
ChestX-ray14 & reinforced reminder & 64 & 64 & 6 & 6 \\
BBBC039 & six-map raw & 200 & 200 & 6 & 6 \\
BBBC039 & six-repeat gray & 200 & 200 & 1 & 6 \\
\bottomrule
\end{tabular}

\smallskip
\parbox{\columnwidth}{\footnotesize Rows are question-answer pairs, and images
are the source-image clusters. Raw marks the base prompt without
the reminder.}
\end{table}

\begin{table*}[t]
\caption{\textbf{Colormap Dependence with Three Answers per Colormap on ChestX-ray14}}
\label{tab:repeated-results}
\centering
\footnotesize
\setlength{\tabcolsep}{4pt}
\begin{tabular}{lrrrr}
\toprule
Model &
\shortstack{r0 drift\\\relax [95\% CI]} &
\shortstack{plug-in MI\\\relax [95\% CI]} &
\shortstack{within-map entropy\\\relax [95\% CI]} &
Holm $p$ \\
\midrule
Claude & 0.257 [0.189, 0.329] & 0.244 [0.180, 0.310] & 0.023 [0.013, 0.035] & 0.0008 \\
Gemini & 0.299 [0.216, 0.388] & 0.171 [0.131, 0.215] & 0.190 [0.147, 0.235] & 0.0020 \\
GPT & 0.151 [0.092, 0.214] & 0.104 [0.066, 0.147] & 0.076 [0.045, 0.113] & 0.0008 \\
Qwen & 0.431 [0.345, 0.515] & 0.236 [0.192, 0.281] & 0.282 [0.232, 0.331] & 0.0008 \\
\bottomrule
\end{tabular}

\smallskip
\parbox{\textwidth}{\footnotesize All quantities except $p$-values are in
nats over 64 source images, with 95\% cluster-bootstrap confidence intervals
(CIs). With 5,000 permutations, 0.0008 is the smallest attainable Holm-adjusted
$p$-value.}
\end{table*}

\subsection{Models and Calls}

We query GPT-5.4, Gemini 3.1 Pro (preview), Claude Sonnet 4.6, and
Qwen3-VL-235B-A22B-Thinking with PNG input, temperature 1.0, low
reasoning effort, and a 4096-token output limit, which Qwen's reasoning exceeds
on 35 of its calls. GPT, Gemini,
and Claude are also asked for the answer schema through structured output, and
Qwen only through the prompt. Temperature 1.0 suits a
measure defined on the sampled answer distribution, and both arms use it, so
sampling variation is measured under the same conditions. The models respond to the
reasoning setting differently: GPT-5.4 and Qwen report reasoning tokens on every
call, Gemini on about 3\% of calls, and Claude on none. Every test compares a model
with itself, and differences between models are reported descriptively. Each
model answers eight conditions, a six-map cell and a six-repeat gray cell for every
panel and a reminder cell for SLAKE and ChestX-ray14
(Table~\ref{tab:validation-matrix-design}). The ChestX-ray14 six-map cell has
three answers per map, and its repeat r0 serves as the one-answer arm. This
gives 5,220 calls per model. Within each cell, a seeded schedule randomizes the
order of the colormaps for each item and rotates it across repeats. Each of
the 20,880 answers is a complete response from the named model that passes the
task schema.

\subsection{Uncertainty Across Images}\label{sec:cluster}

Dataset summaries average item-level scores, and intervals are percentiles of a
cluster bootstrap over source images with 10,000 replicates. Paired contrasts
use the same draw for both arms. For SLAKE, each draw samples 16 images with
replacement and keeps all questions attached to them, and ChestX-ray14 and
BBBC039 resample images directly, because each image carries one question.
Percentile intervals from 16 clusters are approximate, and the confirmatory
test uses ChestX-ray14, where each of the 64 radiographs is its own cluster.
The permutation test uses 5,000 permutations per model and is Holm-corrected
across the four models. The screen reports one interval per model-panel cell
without a multiplicity correction and describes where excess disagreement
appears.

\section{Results}\label{sec:experiments}

We first ask whether the answers depend on the colormap at all and what the
dependence does to them, then how often the matched screen detects it, whether
a repeated instruction or the gray map explains it, and whether the
disagreement it creates can flag wrong answers.

\subsection{All Four Models Depend on the Colormap}

With three answers under each colormap on the 64 chest radiographs, all four
associations exceed the within-item permutation reference after Holm
correction (Table~\ref{tab:repeated-results}), three of them at the smallest
attainable $p$-value. The permutation null means are 0.098, 0.152, 0.069, and
0.196 nats for Claude, Gemini, GPT, and Qwen. Measured against them, Claude's
dependence is 0.146 nats, while Gemini's, GPT's, and Qwen's are 0.019, 0.035,
and 0.040 nats, small but detectable.

Separating the two sources of disagreement changes which model looks most
affected. By disagreement alone, Qwen and Gemini lead, with r0 drift, the orbit
drift of the first repeat, of 0.431 and 0.299 nats against Claude's 0.257, because much of their
disagreement also appears under a single colormap, with within-map entropies
of 0.282 and 0.190 nats. Claude's answers under any one colormap are nearly
fixed, with a within-map entropy of 0.023 nats, yet they differ across
colormaps, so Claude shows the largest dependence once each estimate is
measured against its null. Figure~\ref{fig:teaser} shows this pattern on a
single radiograph: each colormap gives the same answer on all three calls, yet
the colormaps split three to three, an orbit drift of $\log 2=0.693$ nats
against 0 for the six gray repeats. Claude gives a fixed answer under each
colormap that differs across colormaps on 20 of the 64 radiographs, and on all
20 its six gray repeats agree. The disagreement ranking, with Qwen first and GPT last,
holds in each of the three repeats, whose orbit drift varies by at most 0.064
nats per model.

\begin{table}[t]
\caption{\textbf{Accuracy and Abstention under Each Colormap}}
\label{tab:per-map}
\centering
\footnotesize
\setlength{\tabcolsep}{3.2pt}
\begin{tabular}{lrrrrrr}
\toprule
Model & gray & viridis & plasma & inferno & magma & cividis \\
\midrule
\multicolumn{7}{l}{\emph{ChestX-ray14 accuracy}} \\
Claude & 0.56 & 0.49 & 0.49 & 0.58 & 0.56 & 0.53 \\
Gemini & 0.58 & 0.54 & 0.54 & 0.56 & 0.58 & 0.59 \\
GPT & 0.69 & 0.64 & 0.66 & 0.65 & 0.64 & 0.68 \\
Qwen & 0.54 & 0.38 & 0.32 & 0.36 & 0.37 & 0.41 \\
\midrule
\multicolumn{7}{l}{\emph{ChestX-ray14 abstention rate}} \\
Claude & 0.00 & 0.03 & 0.00 & 0.00 & 0.00 & 0.00 \\
Gemini & 0.03 & 0.06 & 0.05 & 0.06 & 0.04 & 0.04 \\
GPT & 0.05 & 0.04 & 0.05 & 0.03 & 0.05 & 0.03 \\
Qwen & 0.08 & 0.34 & 0.38 & 0.31 & 0.28 & 0.27 \\
\midrule
\multicolumn{7}{l}{\emph{SLAKE accuracy}} \\
Claude & 0.80 & 0.74 & 0.74 & 0.78 & 0.80 & 0.82 \\
Gemini & 0.90 & 0.90 & 0.88 & 0.88 & 0.86 & 0.92 \\
GPT & 0.86 & 0.76 & 0.80 & 0.84 & 0.80 & 0.84 \\
Qwen & 0.78 & 0.68 & 0.66 & 0.68 & 0.60 & 0.74 \\
\bottomrule
\end{tabular}

\smallskip
\parbox{\columnwidth}{\footnotesize ChestX-ray14 rates use all three answers under each map on 64
radiographs, and SLAKE accuracy uses the one-answer arm on 50 questions.
Abstentions count as errors.}
\end{table}

\begin{table*}[t]
\caption{\textbf{Matched Drift and Equal-Compute Selective Results}}
\label{tab:primary-results}
\centering
\footnotesize
\setlength{\tabcolsep}{4pt}
\begin{tabular}{llrrrrrrl}
\toprule
Model & Panel & rows/images &
$\widehat D_{\mathrm{orbit}}$ &
$\widehat D_{\mathrm{same}}$ &
$\Delta\widehat D$ [95\% CI] &
orbit AURC &
same AURC &
$\Delta$AURC [95\% CI] \\
\midrule
Claude & BBBC039 & 200/200 & 1.166 & 0.895 &
\textbf{0.271 [0.201, 0.342]} & 0.980 & 0.989 & 0.009 [0.000, 0.029] \\
Claude & ChestX-ray14 & 64/64 & 0.257 & 0.046 &
\textbf{0.211 [0.139, 0.286]} & 0.428 & 0.440 & 0.012 [$-$0.060, 0.086] \\
Claude & SLAKE & 50/16 & 0.140 & 0.009 &
\textbf{0.131 [0.044, 0.224]} & 0.135 & 0.204 &
\textbf{0.069 [0.017, 0.120]} \\
\addlinespace
Gemini & BBBC039 & 200/200 & 1.164 & 1.116 &
0.048 [$-$0.010, 0.107] & 0.985 & 0.982 & $-$0.003 [$-$0.010, 0.000] \\
Gemini & ChestX-ray14 & 64/64 & 0.299 & 0.227 &
0.072 [$-$0.015, 0.164] & 0.384 & 0.386 & 0.003 [$-$0.073, 0.077] \\
Gemini & SLAKE & 50/16 & 0.117 & 0.065 &
0.052 [0.000, 0.113] & 0.035 & 0.070 & 0.035 [$-$0.003, 0.085] \\
\addlinespace
GPT & BBBC039 & 200/200 & 1.613 & 1.606 &
0.007 [$-$0.027, 0.043] & 0.962 & 0.953 & $-$0.008 [$-$0.021, 0.003] \\
GPT & ChestX-ray14 & 64/64 & 0.151 & 0.045 &
\textbf{0.106 [0.051, 0.166]} & 0.265 & 0.302 & 0.037 [$-$0.024, 0.098] \\
GPT & SLAKE & 50/16 & 0.140 & 0.072 &
\textbf{0.068 [0.003, 0.158]} & 0.087 & 0.072 & $-$0.015 [$-$0.109, 0.072] \\
\addlinespace
Qwen & BBBC039 & 200/200 & 1.660 & 1.671 &
$-$0.010 [$-$0.041, 0.020] & 0.956 & 0.954 & $-$0.002 [$-$0.017, 0.015] \\
Qwen & ChestX-ray14 & 64/64 & 0.431 & 0.386 &
0.045 [$-$0.043, 0.138] & 0.454 & 0.422 & $-$0.032 [$-$0.114, 0.048] \\
Qwen & SLAKE & 50/16 & 0.338 & 0.269 &
0.069 [$-$0.044, 0.216] & 0.117 & 0.046 & $-$0.071 [$-$0.137, 0.000] \\
\bottomrule
\end{tabular}

\smallskip
\parbox{\textwidth}{\footnotesize Drift scores are in nats, positive
$\Delta\widehat D$ means that recoloring adds disagreement beyond resampling,
and positive $\Delta$AURC (same-render minus orbit) favors orbit uncertainty.
Intervals are paired per-cell 95\% CIs, and bold ones exclude zero.}
\end{table*}

\subsection{What the Colormap Does to the Answers}\label{sec:effects}

To see what the dependence does, we compare each colored map with gray in the
same three-answer records, taking the majority of the three answers under each
map and none when all three differ. Some colored map overturns the gray
majority on 29 of the 64 radiographs for Claude and on 16 for GPT. Drawing the
six majorities instead from the nine gray answers of each radiograph, with
replacement, overturns it on only 3.9 [2, 6] and 4.6 [2, 7] radiographs (mean
and central 95\% range of 2,000 draws), so these changes follow the
display. They include 17 of the 37 radiographs on which Claude's gray majority matches
the label and 9 of the 45 for GPT, against 3.2 [1, 5] and 2.9 [1, 5] under resampling.
Gemini's 14 changes lie within its resampling range of 7 to 15.

The models also fail in different directions (Table~\ref{tab:per-map}).
Averaged over the five colored maps, Qwen's abstention rate on ChestX-ray14
rises by 0.236 [0.172, 0.302] over gray, 83 of its 91 changed majorities are
abstentions, and its accuracy falls by 0.168 [0.106, 0.234] on ChestX-ray14 and
by 0.108 [0.049, 0.173] on SLAKE. An abstention is not a wrong diagnosis,
but it withholds an answer that Qwen gives under gray, so the display decides
whether it answers at all. GPT keeps answering, but its yes rate on
ChestX-ray14 falls by 0.130 [0.066, 0.201]. Claude rarely abstains and becomes
more consistent under color. On ChestX-ray14 its within-map entropy is 0.014
nats under the colored maps against 0.070 under gray, a difference of $-0.056$
[$-0.111$, $-0.008$], so its dependence there is a shift between consistent
answers. Gemini becomes noisier under color, with within-map entropy of 0.202
nats against 0.127, a difference of 0.076 [0.009, 0.139]. Because the screen
measures resampling under gray, these two cases show how the reference affects
it: Claude's excess arises despite quieter colored renderings, and Gemini's
noisier colored renderings would inflate its contrast, which nonetheless stays
unresolved.

\subsection{The Matched Screen}

With one answer per colormap, recoloring adds disagreement beyond same-render
variation in 5 of 12 cells under per-cell 95\% intervals, namely every Claude
panel and GPT on ChestX-ray14 and SLAKE (Table~\ref{tab:primary-results}),
with Gemini on SLAKE at the boundary, 0.052 [0.000, 0.113]. No
interval lies entirely below zero, so in no cell did asking again produce
reliably more disagreement than recoloring. The control matters most where a
model disagrees with itself regardless of the display. On BBBC039, GPT and Qwen
return mostly different exact counts for six repeats of the identical gray
image, with same-render drift of 1.606 and 1.671 nats against a maximum of
1.792. A test without the control would rank these as the most
colormap-sensitive cells, whereas the matched contrast detects no excess
beyond resampling. Near this ceiling the entropy scale has little room, so we
also read the counts as numbers, with the same result.
On the 197 to 198 images where both arms return at least two numeric counts,
the mean absolute deviation of the counts from their median exceeds that of the
gray repeats by 2.36 [1.85, 2.87] nuclei for Claude, and the intervals for
Gemini, GPT, and Qwen all contain zero. In the other
direction, the screen can miss dependence. For Gemini and Qwen on ChestX-ray14 the
matched contrast is unresolved, yet the three-answer test detects dependence,
so a silent screen is no evidence of invariance.

\subsection{A Second Instruction and the Gray Map}

Repeating the instruction to ignore color changes drift detectably in two of
eight cells, in opposite directions. It lowers drift for Claude on SLAKE by
0.048 [0.015, 0.078] nats and raises it for GPT on SLAKE by 0.044 [0.011,
0.079] nats. Accuracy rises for Qwen on ChestX-ray14 by 0.055 [0.018, 0.094],
and the other seven accuracy intervals cross zero. Gray leverage is resolved in
2 of 12 cells, for Claude on BBBC039 with 0.043 [0.017, 0.069] nats and for GPT
on ChestX-ray14 with 0.034 [0.001, 0.072] nats. In the other
ten cells gray leverage is unresolved and within 0.015 nats of zero, so there is
no sign of a split between gray and the colored maps. Figure~\ref{fig:teaser} shows the same picture for a single
radiograph, where plasma and cividis agree with gray while viridis, inferno,
and magma do not.

\subsection{Recoloring as an Error Signal}

Table~\ref{tab:primary-results} also compares the two disagreement signals as
detectors of wrong answers with the same number of calls. Recoloring helps in
one cell, Claude on SLAKE, where asking again carries almost no signal: its six
gray repeats disagree on one of 50 rows (0.009 nats), and there orbit
uncertainty reaches an
AURC of 0.135 compared with 0.204 for same-render uncertainty, a difference of
0.069 [0.017, 0.120], and the same-render AURC is 0.204 under all six omission
rotations. In the other eleven cells neither signal is reliably better, although the
intervals for Qwen on SLAKE and for Claude and Gemini on BBBC039 end at zero. In the
eight medical cells the model's stated confidence, from a single call, has a
lower AURC than orbit uncertainty, although none of these differences is
resolved. Recoloring is therefore best read as a diagnostic of colormap
dependence that added an error signal in one cell, rather than as a general
uncertainty estimate.

\section{Discussion}\label{sec:discussion}

MCD needs only the model's answers, so colormap invariance can be tested on any
question whose answer should not depend on the display, including panels whose
labels are noisy or missing, such as the NLP-derived ChestX-ray14 labels. Where labels exist, they show that the dependence matters: under color Qwen
loses accuracy on both medical panels, mostly by abstaining, and for Claude and
GPT the display overturns label-matching gray answers far more often than
resampling does, even where average accuracy barely moves.

Three practical lessons follow. First, the colormap belongs in the description
of any VLM evaluation on scalar images, next to windowing and resolution,
because it can change the model's answers. Second, a perturbation test for a
stochastic model needs a resampling control with the same number of calls.
Without it, the BBBC039 counts of GPT and Qwen would read as extreme colormap
sensitivity, and with it the excess over resampling is unresolved, on the
entropy scale and on the count scale alike. Third, an instruction does not
prevent the dependence, and repeating it did not reliably reduce drift.

The two steps complement each other. Twelve calls per item, six per arm,
screen for excess disagreement and resolve it in five of twelve cells here,
and eighteen calls per item detected dependence in the two ChestX-ray14 cells
where the screen was silent.

\noindent\textbf{What the colormap changes.} The six maps share the order of
intensities but differ in hue, luminance profile, and local contrast. Two
observations argue
against simpler accounts. Removing gray, the only achromatic map, changes
disagreement no more detectably than removing a chromatic map in ten of twelve
cells, and on
ChestX-ray14 Claude answers each colored rendering more consistently than the
gray one, so its changes across maps follow the display. Vision models are known to rely on
cues outside the intended signal, such as
texture~\cite{geirhos2019texture,geirhos2020shortcut}. Benign changes of the
input, such as a one-pixel shift or a small resize, and luminance-matched
palettes would separate colormap-specific sensitivity from general input
sensitivity and the role of hue from that of lightness. Because users choose
the colormap freely, dependence on it matters even if part of it reflects
general input sensitivity.

\noindent\textbf{Scope and outlook.} The permutation test establishes
dependence for all four models on the 64 chest radiographs, and on SLAKE and
BBBC039 the matched screen detects excess disagreement for Claude on both
panels and for GPT on SLAKE. MCD measures behavior: it shows that the answers
depend on the display, not why, and it does not by itself measure clinical
harm, which would need expert-adjudicated labels. Three answers per map on
SLAKE, where the models are more accurate, and a broader survey across tasks
and modalities would show how large and how common the dependence is. We query rendered images with a
fixed prompt and settings. Medical VLMs, systems that read raw pixel arrays or
adjust windowing, and agents that call image tools can be tested with the same
protocol, and inverted gray, windowing, modality presets such as Hot Iron, and
single-hue fluorescence lookup tables are natural additions to the set of
transformations. The release, linked from the project page at
\href{https://siddvoh.com/mutual-colormap-dependence/}{\mbox{\nolinkurl{siddvoh.com/mutual-colormap-dependence}}},
contains the experiment plan with its prompts and settings, the three item
panels, all 20,880 answers, and the analysis code that rebuilds every result
from these answers.

\section{Conclusion}\label{sec:conclusion}

A colormap is part of what a VLM sees when it reads a biomedical image, and it
can change the answer. Mutual Colormap Dependence turns colormap invariance
into a quantity that can be measured from a model's answers alone, and comparing
recoloring with a resampling control of the same size keeps ordinary sampling
from being read as an effect of the colormap. Applied to four frontier VLMs, the test finds
that all four change their answers about chest radiographs with the colormap,
even when told to ignore it, that these
changes overturn answers that agree with the labels for Claude and GPT and lower
Qwen's accuracy, and that a repeated instruction does not reliably help. We hope
that MCD and its matched control become a routine check whenever a VLM is
evaluated on scalar biomedical images.

\paragraph{Acknowledgments.}
This work was supported in part by U.S. NIH grant R35GM158094.
The Gemini calls in this study were supported by a Gemini Academic Program award of Google Cloud credits from Google.
Google had no role in study design, execution, analysis or reporting, and did not review or influence this work.
Generative AI tools assisted in preparing this work, and the authors reviewed all content and take full responsibility for it.

{
    \small\raggedright
    \bibliographystyle{ieeenat_fullname}
    \bibliography{refs}
}

\end{document}